\documentclass[sigconf]{acmart}

\setcopyright{none}              
\renewcommand\footnotetextcopyrightpermission[1]{} 

\usepackage{algorithm}
\usepackage{algpseudocode}
\usepackage{graphicx}
\usepackage{subcaption}
\usepackage{textcomp}
\usepackage{xcolor}
\usepackage{tabularx}
\usepackage{multirow}
\usepackage{float}
\usepackage{multicol}
\usepackage{booktabs}
\usepackage{makecell}
\usepackage{balance}

\usepackage{pifont}
\newcommand{\cmark}{\ding{51}}
\newcommand{\xmark}{\ding{55}}
\acmConference[]{}{}{}
\title{Arisca: A Parameterized Symbolic Algebra Framework for Arithmetic Circuit Verification}

\author{Kezhi Li}

\affiliation{
    \institution{The Chinese University of Hong Kong}
    \city{Hong Kong}
    \country{China}
}

\author{Min Li}
\affiliation{
    \institution{Southeast University}
    \city{Nanjing}
    \country{China}
}

\author{Qiang Xu}
\affiliation{
    \institution{The Chinese University of Hong Kong}
    \city{Hong Kong}
    \country{China}
}

\begin{abstract}
Formal verification of highly optimized arithmetic circuits at the gate-level remains a significant challenge due to the state space explosion problem. Although Symbolic Computer Algebra (SCA) offers a scalable theoretical foundation by modeling circuits as multivariate polynomials, practical implementations frequently suffer from the explosion of the size of intermediate polynomials. State-of-the-art SCA tools typically rely on fixed heuristics and restrict their application to standard multipliers. A fixed heuristic is insufficient for structurally diverse arithmetic circuits, as it often fails to generalize across all cases. In this paper, we introduce Arisca, an open-source parameterized verification framework for \textbf{Ari}thmetic circuits using \textbf{S}ymbolic \textbf{C}omputer \textbf{A}lgebra. 
Arisca establishes a generalized parameter space that unifies previously isolated state-of-the-art (SOTA) techniques as specific configurations within a broader algebraic reduction theory. To fundamentally transplant and elevate previous methods, we propose several algorithmic improvements, such as an HA-preserving extraction strategy, density-based vanishing detection, and conservative polynomial size estimation. 
In addition, Arisca expands the verification scope to encompass general arithmetic circuits with any combination of addition and multiplication, such as multiply-accumulators and dot-product units.
Extensive evaluations demonstrate that Arisca achieves SOTA performance in a comprehensive suite of multiplier benchmarks and a diverse array of practical arithmetic cases.
\end{abstract}


\begin{document}

\maketitle

\section{Introduction}
\label{sec:intro}

Arithmetic circuits are fundamental components of modern microprocessors and hardware accelerators \cite{jouppi2023TPUa, tirumala2024nvidia}. As demonstrated by the 1994 Pentium bug, the financial costs of arithmetic logic errors can be catastrophic \cite{price1995Pentium}. Therefore, a sufficient formal verification of the circuit has become an indispensable step in the Electronic Design Automation (EDA) flow. To ensure that logic synthesis tools do not introduce functional bugs during optimizations, the gate-level implementation---which serves as the final deliverable of the front-end flow---must be thoroughly verified. However, fully automated verification of highly optimized arithmetic circuits, such as multipliers, at the gate level (e.g., And-Inverter Graphs, AIGs) remains one of the most challenging tasks in formal verification due to their immense structural complexity and the inherent state space explosion problem.

To tackle this challenge, researchers have explored various approaches. Traditional equivalence checking methods that use Binary Decision Diagrams (BDDs) often suffer memory explosion when applied to integer multipliers \cite{bryant1991Complexity, kumar2023Formal}. Similarly, while approaches based on Boolean Satisfiability (SAT)  solvers \cite{goldberg2001using, kaufmann2019Verifying} and Satisfiability Modulo Theories (SMT) solvers \cite{demoura2008Z3} excel at handling control logic, they frequently encounter exponential runtime when confronted with deeply optimized datapath architectures. Furthermore, emerging methods based on theorem proving, such as the VeSCMul framework utilizing SC-Rewriting \cite{temel2020Automated, temel2021Sound, temel2024VeSCMul, temel2024Formal}, have gained attention and shown promising results. However, these techniques are dependent on the preservation of hierarchical high-level boundaries and structural information within the circuit, rendering them inapplicable to completely flattened gate-level netlists. In contrast, verification techniques based on Symbolic Computer Algebra (SCA) offer a scalable alternative by defining the equivalence checking task as an ideal membership problem \cite{sayed-ahmed2016Formal}. Specifically, SCA models both the gate-level netlist and the arithmetic specification as multivariate polynomials over a ring. By applying polynomial reduction---typically through backward rewriting rooted in Gr\"obner basis theory---SCA captures word-level arithmetic signatures, effectively bypassing the bit-level state space explosion. 

Despite its theoretical soundness, the direct application of SCA to large, highly optimized arithmetic circuits in practice can lead to an explosion of the intermediate polynomial size, posing a significant challenge to scalability. To make this approach practical for such designs, various optimizations have recently been proposed. For example, structural techniques such as XOR-based slicing and final adder replacement have been introduced to streamline algebraic reasoning \cite{kaufmann2023Improving, kaufmann2019Verifying}. Integration of dual variables by \cite{kaufmann2022Adding} further mitigates polynomial size blowup.  Meanwhile, \cite{mahzoon2018PolyCleaner, mahzoon2022RevSCA20} leverage reverse engineering to extract internal structures such as Full Adders (FA) and Half Adders (HA), which not only reduces the number of intermediate variables but is also crucial to identify and remove ``vanishing monomials'', a key factor contributing to the explosion of polynomial size . Furthermore, dynamic heuristics that adjust variable substitution orders and signal phases have proven effective in strictly controlling polynomial growth during the rewriting process \cite{konrad2024Symbolic}.

Despite these advances, the industrial adoption of SCA-based verification remains limited. A key reason is the restricted accessibility of state-of-the-art (SOTA) implementations: many leading verification engines are released solely as closed-source binaries, typically supporting only fixed $n \times n$ multipliers. This limited accessibility leads to two interrelated challenges. First, it restricts the community's ability to extend SCA to other ubiquitous arithmetic structures---such as multiply-accumulators (MACs) and vector dot-product units---that SCA could support without much theoretical enhancement. Second, and more fundamentally, it prevents systematic combination and optimization of the underlying heuristics. Arithmetic circuits can exhibit significantly different architectures when implementing the same function \cite{sklansky1960conditional, wallace2006suggestion, dadda1965some, brent1982regular}. For example, integer multipliers have booth coding with different radixes to encode their partial products. Therefore, a single fixed heuristic is unlikely to generalize to all possible designs. Moreover, many existing optimizations remain in an ``under-explored'' state, applied in isolation without the parameterization needed to combine or adapt them for diverse circuit topologies. Consequently, these methods have not yet reached their full potential in applicability and performance, leaving a significant gap between academic research and industrial practice.

To overcome these challenges, this paper introduces Arisca, an open-source generalized verification framework for \textbf{Ari}thmetic circuits based on \textbf{S}ymbolic \textbf{C}omputer \textbf{A}lgebra. At its core, Arisca transitions from a rigid toolchain to a generalized verification engine. Rather than merely assembling existing methods into a single tool, we introduce a parameterized space where previously isolated heuristics become specific configurations of a unified reduction theory. To enable this parameterization and refine previous techniques, we propose several fundamental algorithmic improvements, such as an HA-preserving extraction strategy, density-based vanishing detection, and conservative polynomial size estimations. Furthermore, Arisca implements a robust parser to solve general arithmetic circuits with any combination of addition and multiplication. This effectively expands the verification scope of previous tools to natively support a broader class of realistic designs. Extensive evaluations demonstrate that Arisca achieves SOTA performance in a comprehensive suite of multiplier benchmarks and a diverse array of practical arithmetic use cases. An anonymous version of the source code is available for peer review at \url{https://anonymous.4open.science/r/arisca-C4E8}. The complete repository will be open-source upon acceptance.

\section{Preliminaries}
\label{sec:preliminary}
In this section, we briefly define the algebraic framework used for the formal verification of general arithmetic circuits, drawing on the principles of SCA and the basis theory of Gr\"obner.

\subsection{Algebraic Modeling of Arithmetic Circuits}
We represent a gate-level arithmetic circuit as a set of Boolean variables $X = \{x_1, \dots, x_n\}$ corresponding to primary inputs, and $Z = \{z_1, \dots, z_m\}$ corresponding to internal signals and primary outputs. Each logic gate in the circuit is associated with a polynomial $g_i \in \mathbb{Q}[X, Z]$ that characterizes its logical behavior. For example, an AND gate $z_k = \text{AND}(z_i, z_j)$ is mapped to $g_k = z_k - z_i z_j$.

The set of all gate polynomials $G = \{g_1, \dots, g_{|Z|}\}$ generates an ideal $J = \langle G \rangle \subseteq \mathbb{Q}[X, Z]$. A key property exploited in SCA-based verification is that if the variables in $Z$ are ordered according to a reverse topological sorting of the circuit netlist, the set $G$ naturally forms a Gr\"obner basis for the ideal $J$ in the order of lexicographic terms. This structure ensures that the reduction of any polynomial modulo $G$ produces a unique remainder.

\subsection{Word-Level Specification and Ideal Membership}
\label{subsec:wdspec}
The functional correctness of an arithmetic circuit is defined by a specification polynomial $Sp$. To construct $Sp$, an $m$-bit signal vector $Y = \{y_0, \dots, y_{m-1}\}$ is mapped to its value at the word-level using an evaluation function $W(Y)$. Depending on the data type, $W(Y)$ is defined as follows:

\begin{itemize}
    \item Unsigned: $W(Y) = \sum_{j=0}^{m-1} 2^j y_j$
    \item Signed: $W(Y) = -2^{m-1} y_{m-1} + \sum_{j=0}^{m-2} 2^j y_j$
\end{itemize}
Let $I \subseteq X$ be the set of input vectors and $O \subseteq Z$ be the $k$-bit output vector. The specification is expressed as follows.
$$Sp = W(O) - \mathcal{F}(W(I)),$$
where $\mathcal{F}$ represents the intended arithmetic function at the word-level. The verification problem is then reduced to an ideal membership test: the circuit is formally verified if and only if $Sp \in J$. In practice, this is checked by computing the remainder of $Sp$ modulo $G$. If $Sp \xrightarrow{G}_+0$, the circuit matches its specification.

\subsection{Verification via Symbolic Rewriting}
The computation of the remainder is performed through a process known as backward rewriting. Starting from $Sp$, we successively substitute each variable $z_i \in Z$ with its representation of polynomials in reverse topological order. Although the Gr\"obner basis property theoretically guarantees a zero remainder for functionally correct circuits, the practical execution of backward rewriting is often computationally prohibitive. Intermediate polynomials can easily suffer from an exponential increase in size due to the accumulation of monomials. Effectively managing this symbolic complexity during the reduction process remains the primary hurdle in scaling algebraic verification for large arithmetic designs.

\section{ARISCA}
\label{sec:method}
In this section, we detail the core methodologies implemented in Arisca. The tool takes an AIG file as input. Without an explicit specification, it defaults to $n \times n$ multipliers. Arisca evaluates whether the circuit is correct according to the specification. If the circuit is incorrect, the tool additionally outputs a non-zero residue polynomial. Section \ref{subsec:spec} introduces Arisca's generalized framework for handling arithmetic circuit specifications beyond standard multipliers. Section \ref{subsec:re} presents the reverse engineering techniques for circuit preprocessing, followed by Section \ref{subsec:atomic}, which explains the algorithm for extracting atomic blocks from the netlist. To optimize the polynomial reduction process, Section \ref{subsec:order} introduces a dynamic rewriting heuristic and proposes a phase inversion method. Finally, Section \ref{subsec:portfolio} presents a portfolio version of Arisca that integrates all these techniques.

\subsection{Specification Parsing}
\label{subsec:spec}
To extend the verification ability beyond standard multipliers, Arisca introduces a flexible specification parser that allows users to define custom arithmetic functions $\mathcal{F}$ introduced in Section \ref{subsec:wdspec}. Instead of hardcoding the specification polynomial, users can provide a mathematical expression consisting of variables, addition (\texttt{+}), multiplication (\texttt{*}), and parentheses.

\begin{figure}[tbp]
    \centering
    \includegraphics[width=\linewidth]{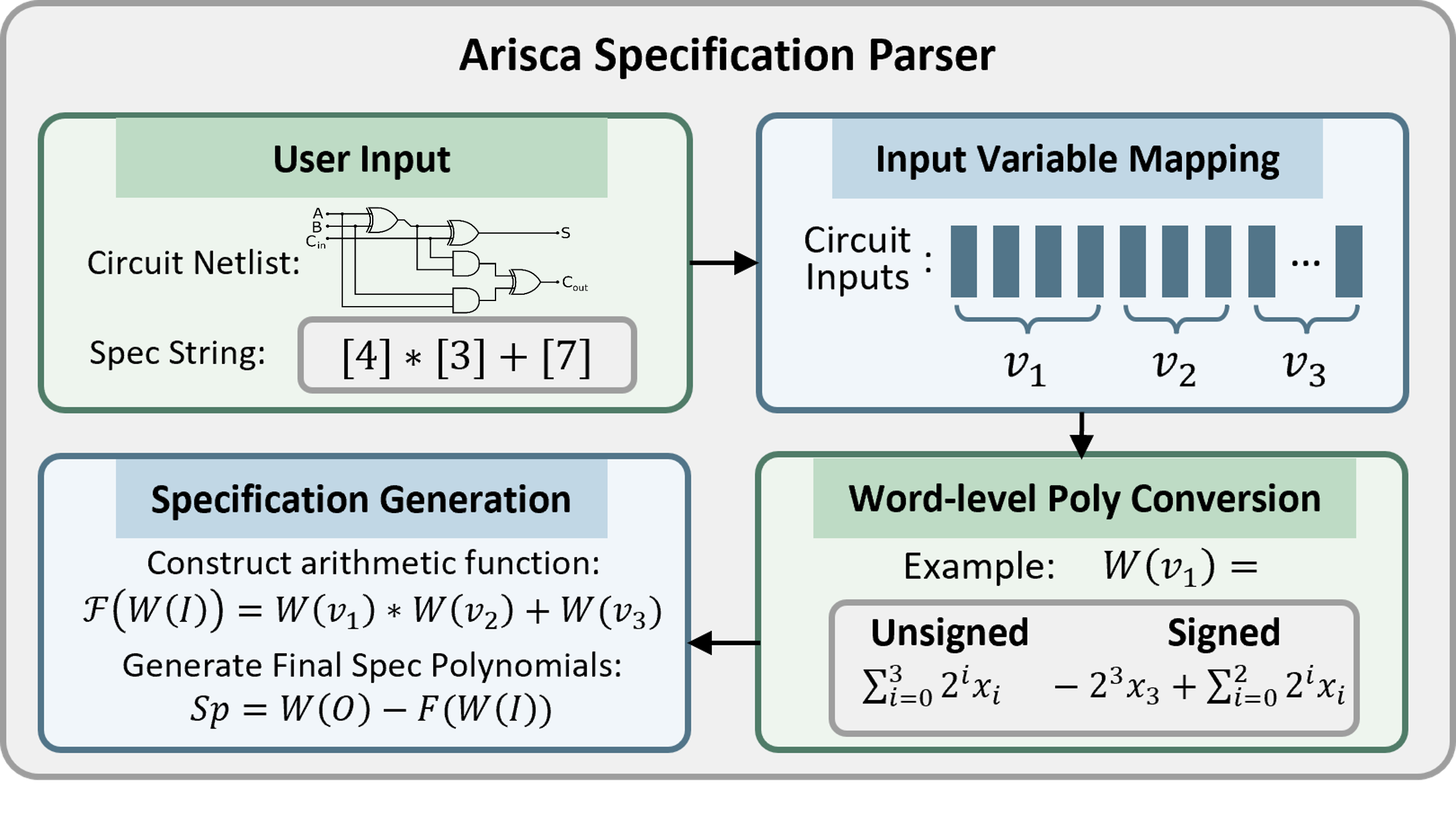}
    \caption{Overview of the specification parsing and polynomial construction workflow in Arisca.}
    \label{fig:parser}
    \vspace{-2mm}
\end{figure}

To support varying input sizes, a variable is denoted by the syntax \texttt{[n]}, where $n \in \mathbb{Z}^+$ specifies the bit-width of the operand. As shown in Fig.~\ref{fig:parser}, during the initial parsing phase, Arisca extracts all variables from the user-provided expression and sequentially maps them to the primary inputs of the given netlist in their order of appearance. To provide further flexibility, Arisca also supports an explicit offset syntax \texttt{[n:offset]}, which explicitly maps the variable to $n$ bits starting from the specified \texttt{offset} index. To guarantee structural consistency, the parser asserts that the maximum input boundary accessed by these variables exactly matches the total number of primary inputs in the circuit. If a mismatch is detected, Arisca raises an error and aborts the execution.

Once the variable mapping is established, Arisca constructs the word-level representation for each input operand. Depending on a user-configured flag \texttt{unsigned} or \texttt{signed}, the tool applies the corresponding evaluation function $W(\cdot)$ defined in Section \ref{subsec:wdspec} to each input vector. The parser then evaluates the abstract syntax tree of the user-defined expression using these word-level operands, constructing the target arithmetic function $\mathcal{F}(W(I))$. Similarly, a word-level evaluation of the primary outputs of the circuit, $W(O)$, is generated. The final specification polynomial is thus formulated as $Sp = W(O) - \mathcal{F}(W(I))$, serving as the starting point for the subsequent backward rewriting process.

Consider a typical MAC unit. The user can simply provide the expression \texttt{[4] * [3] + [7]}. Arisca parses this expression and maps the first 4 bits, the subsequent 3 bits, and the final 7 bits of the circuit's primary inputs to the three operands, respectively. It then multiplies the word-level polynomials of the first two operands, adds the third to form the specification of the input $\mathcal{F}(W(I))$. Finally, it subtracts the resultant $\mathcal{F}(W(I))$ from the evaluated output $W(O)$ to construct the complete specification polynomial $Sp$.

Although the current implementation of the parser is restricted to addition and multiplication operations, the underlying SCA framework is inherently applicable to a much broader class of arithmetic circuits \cite{mahzoon2022Formal, drechsler2023Divide, konrad2025Divider, lv2013Efficient}. Extending Arisca to natively support a wider range of arithmetic components, paired with dedicated algebraic reduction heuristics, remains a part of our future work.

\subsection{Reverse Engineering}
\label{subsec:re}
In standard design flows, logic synthesis flattens the hierarchical structure of arithmetic circuits, introducing intermediate variables that make direct symbolic backward rewriting inefficient. To address this, reverse engineering is employed as a preprocessing step to abstract the flattened netlist back into functionally equivalent macroscopic blocks, such as FAs and HAs. As demonstrated in previous studies \cite{mahzoon2019RevSCA, mahzoon2022RevSCA20}, the reconvergent outputs of HAs are a major source of vanishing monomials, contributing to the polynomial explosion problem. Therefore, ensuring robust extraction of HAs is critical for scalable algebraic verification.

\begin{figure}[tbp]
    \centering
    \includegraphics[width=0.9\linewidth]{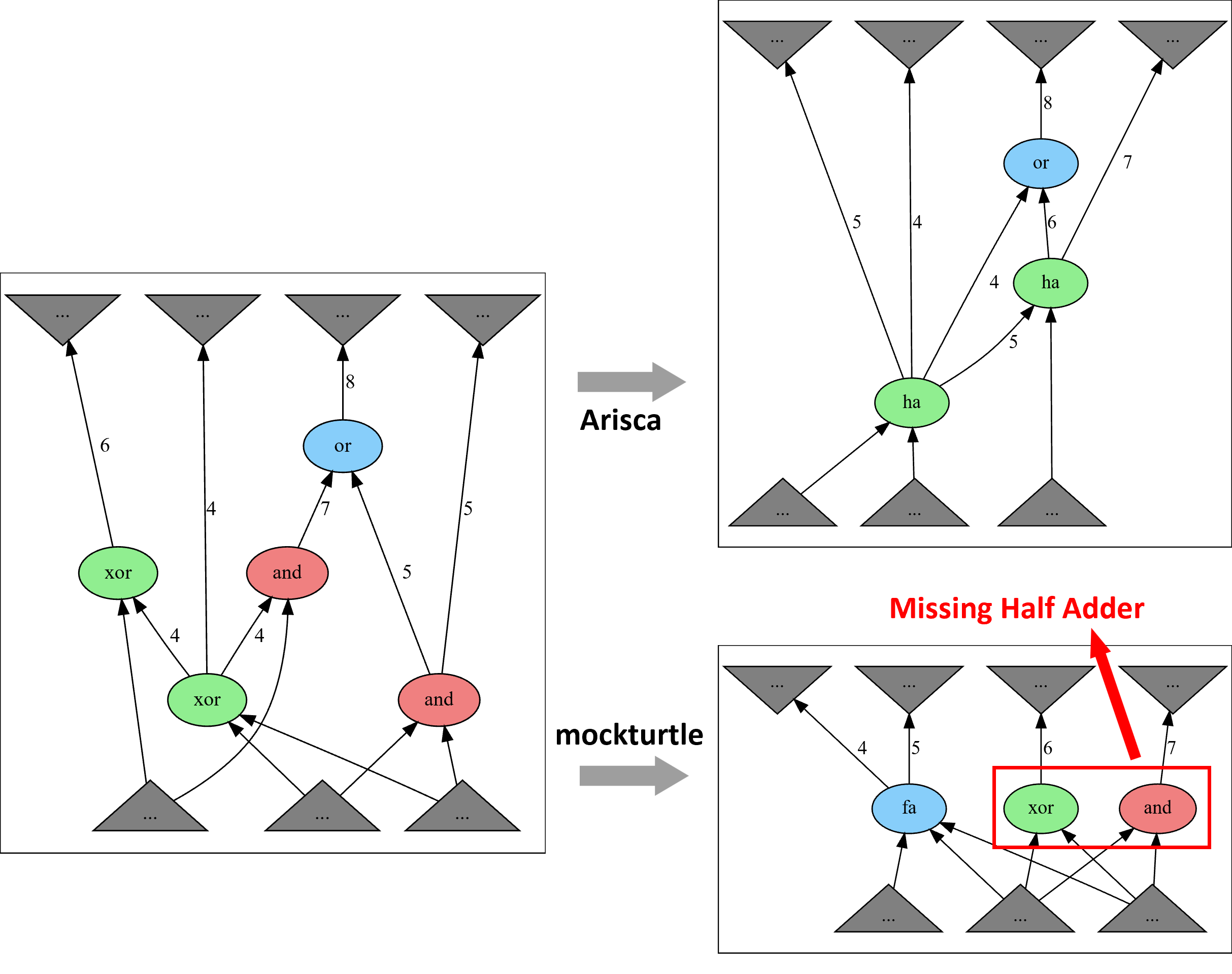}
    \caption{An FA candidate containing internal nodes with external fanouts. While a greedy FA-first strategy forces logic duplication and obscures the constituent HA, Arisca's conservative approach successfully preserves and extracts it.}
    \label{fig:extract}
    \vspace{-3mm}
\end{figure}

The open-source library \texttt{mockturtle} is frequently used for cut enumeration and block extraction in previous SCA tools \cite{mahzoon2022RevSCA20}. However, its default extraction algorithm employs an FA-first mapping priority. Although this approach guarantees functional equivalence, it can miss smaller constituent components such as HAs when processing netlists with dense logic sharing. To illustrate this, consider the sub-circuit depicted in Fig.~\ref{fig:extract}. The left sub-circuit can be mapped as an FA, with signal 6 generating the sum and signal 8 computing the carry. However, internal nodes within this FA candidate, signal 4 and signal 5, possess additional fanouts driving external logic. When the algorithm maps this entire block as an FA, it must duplicate the predecessor logic cones of signals 4 and 5 to maintain functional equivalence for these external fanouts. As shown in the lower-right part of Fig.~\ref{fig:extract}, the duplicate logic that inherently forms an HA is left unmapped. 

To tackle this problem, Arisca implements its own cut enumeration and gate mapping engines. To preserve the constituent components, it adopts an HA-preserving strategy. It rejects to map an FA candidate if any of its internal nodes have external fanouts. Instead, it goes back to mapping the constituent components, thereby extracting the underlying HA. This conservative mapping strategy avoids logic duplication and preserves the HA structures required to identify vanishing monomials. To empirically validate this approach, we compared the average number of HAs and FAs extracted from the Akoi benchmark using \texttt{mockturtle} and Arisca. As shown in Table~\ref{tab:ha_fa_comparison}, Arisca successfully extracts a higher number of HAs on average by avoiding aggressive FA-first mapping. Furthermore, the total combined number of extracted adders remains comparable between the two methods, demonstrating that our HA-preserving strategy does not sacrifice the overall level of macroscopic abstraction.

Beyond multi-fanout gates like HA and FA, Arisca supports common single-output gates such as XOR and MAJ. It allows users to specify a customized extraction sequence. For example, given the argument ``\texttt{-e adder,xor,maj}'', Arisca sequentially extracts HAs and FAs, followed by XOR and MAJ gates from the circuit. Note that this extraction order directly affects the resulting circuit structure due to the phase ordering problem.

\begin{table}[tbp]
  \centering 
  \caption{Comparison on Average Extracted Adders.}
  \label{tab:ha_fa_comparison}
  \setlength{\tabcolsep}{4mm} 
  \begin{tabular}{ccc}
    \toprule
    \textbf{Metric} & \textbf{\texttt{mockturtle}} & \textbf{Arisca} \\
    \midrule
    HAs & 1021.95 & 1124.98 \\
    FAs & 2394.66 & 2343.14 \\
    \midrule
    Total & 3416.61 & 3468.12 \\
    \bottomrule
  \end{tabular}
\end{table}

\subsection{Atomic Block Extraction}
\label{subsec:atomic}

Beyond the identification of FAs and HAs described in Section \ref{subsec:re}, encapsulating arbitrary fanout-free logic cones into independent atomic blocks further reduces the number of intermediate variables during polynomial reduction \cite{farahmandi2015Groebnera}. The algorithm \ref{alg:extraction} outlines our integrated approach to extracting these logic cones and simultaneously identifying candidates for the removal of early vanishing monomials.

The extraction process begins by defining the structural boundaries within the netlist. We designate Primary Inputs (PIs), the outputs of previously mapped FAs and HAs, and any nodes with multiple fanouts as termination nodes $\mathcal{T}$. For each multi-fanout node, the algorithm traces backward through the netlist until it encounters termination nodes, thereby constructing an initial set of independent logic cones (Lines 4-7). 

Once the initial cones are formed, Arisca aims to identify converging logic structures for the early removal of vanishing monomials. Previous approaches \cite{mahzoon2018PolyCleaner, mahzoon2022RevSCA20} attempt to perform early elimination for every converging logic structure of HAs. This requires the identification of thousands of reconvergent path pairs. To avoid this computational overhead, Arisca employs a density-based heuristic evaluated directly on the input of the logic block.

However, evaluating this density metric on small, fragmented initial cones may not capture the driving HA pairs hidden behind adjacent logic. To expose these pairs and broaden the receptive field, Arisca introduces a speculative cone extension step. The algorithm iterates through each initial cone $C$. If the cone meets a minimum size threshold $|C| \ge 3$, it dynamically absorbs smaller adjacent cones ($|C'| < 5$) by queueing and evaluating their inputs (Lines 9-19). This temporary expansion consolidates fragmented logic, providing richer context for identifying HA pairs.

Following the extension, Arisca evaluates the finalized cone. We count the number of pairs of HA outputs present among the cone inputs (denoted $P$). If condition $(2 \times P) / |\text{inputs of } C| > r / 10$ is satisfied, where $r \in [0, 10]$ is a user-defined sensitivity parameter, the block is classified as a vanishing cone and marked for early removal during the subsequent rewriting phase. If the extended cone does not meet this density threshold, the algorithm restores the cone to its original state (Lines 20-25). Using this density metric within a unified extraction pass, Arisca bypasses the exhaustive search for reconvergent paths.

\begin{algorithm}[tbp]
\caption{Atomic Block Extraction and Vanishing Detection}
\label{alg:extraction}
\begin{algorithmic}[1]
\renewcommand{\algorithmicrequire}{\textbf{Input:}}
\renewcommand{\algorithmicensure}{\textbf{Output:}}

\Require Flattened netlist $\mathcal{N}$, Extracted HAs/FAs $\mathcal{A}$, Sensitivity parameter $r \in [0, 10]$ 
\Ensure Set of extracted atomic blocks $\mathcal{B}$

\State $\mathcal{T} \leftarrow$ Primary Inputs $\cup$ Outputs of $\mathcal{A} \cup$ Multi-fanout nodes \Comment{Termination nodes}
\State $\mathcal{S} \leftarrow$ Multi-fanout nodes \Comment{Start nodes}
\State $\mathcal{C}, \mathcal{B} \leftarrow \emptyset$

\For{each $s \in \mathcal{S}$}
    \State $C \leftarrow \text{Traverse backward from } s \text{ until reaching nodes in } \mathcal{T}$
    \State $\mathcal{C} \leftarrow \mathcal{C} \cup \{C\}$
\EndFor

\For{each cone $C \in \mathcal{C}$}
    \State $C_o \leftarrow C$ \Comment{Store the original cone}
    \If{$|C| \ge 3$} \Comment{Cone extension}
        \State $Q \leftarrow \text{queue containing the inputs of } C$ 
        \While{$Q$ is not empty}
            \State $v \leftarrow Q.pop()$
            \If{$v$ is the root of a cone $C' \in \mathcal{C}$ \textbf{and} $|C'| < 5$}
                \State Merge $C'$ into $C$
                \State Push the inputs of $C'$ into $Q$
            \EndIf
        \EndWhile
    \EndIf
    \State $P \leftarrow \text{Number of HA output pairs in the inputs of } C$
    \If{$(2 \times P) / |\text{inputs of } C| > r / 10$}
        \State Mark $C$ as a vanishing cone
    \Else
        \State $C \leftarrow C_o$
    \EndIf
    \State $\mathcal{B} \leftarrow \mathcal{B} \cup \{C\}$
\EndFor

\State \textbf{return} $\mathcal{B}$
\end{algorithmic}
\end{algorithm}

\subsection{Dynamic Rewriting Ordering \& Variable Phase Inversion}
\label{subsec:order}

Relying solely on static reverse topological orders is often insufficient to prevent intermediate polynomial explosion during backward rewriting. To provide operational flexibility, Arisca exposes a configurable parameter \texttt{mode}. It supports static topological traversals, specifically Breadth-First Search (BFS) and Depth-First Search (DFS). Furthermore, Arisca incorporates an optional dynamic look-ahead rewriting order, inspired by \cite{mahzoon2020Formal, konrad2024Symbolic}. The Algorithm \ref{alg:ordering} details this dynamic substitution process.

The core of dynamic ordering lies in estimating the polynomial size expansion before committing to a substitution. During each iteration, the algorithm maintains a set of ready candidates $\mathcal{V}_{ready}$ that meet the reverse topological constraints. Previous work commonly ranks candidates solely on their frequency $n_v$ in the current polynomial $\mathcal{P}$. This aggressive strategy ignores the substituted polynomial size $|\mathcal{P}_v|$, risking explosive growth when $|\mathcal{P}_v|$ is large. In reality, replacing a variable $v$ expands the overall polynomial by at most $(n_v - 1) \times |\mathcal{P}_v|$. The ranking of candidates proportional to this upper bound by $n_v \times |\mathcal{P}_v|$ yields a conservative ordering that effectively restricts size inflation. To accommodate diverse circuit structures, Arisca introduces a configurable flag to toggle between aggressive and conservative estimation heuristics. The candidates are then sorted in ascending order by the selected estimation value multiplied by a penalty factor that is initialized to 1 (Lines 4-8).

Since sorted candidates may still cause polynomial size explosions, Arisca employs a look-ahead rewriting scheme (Lines 9-23). The algorithm temporarily substitutes the candidate and computes the growth ratio of the polynomial size $\Delta$. This ratio is evaluated against a user-defined explosion threshold $\theta_{max}$. If $\Delta \le \theta_{max}$, the substitution is accepted, and the algorithm proceeds to the next iteration. In contrast, if the threshold is exceeded, the substitution is reversed. In addition, the penalty factor for the rejected variable $p(v)$ is doubled (Line 18) so that these explosive variables will not dominate the queue \cite{konrad2024Symbolic}. The algorithm then proceeds to the next candidate. If all ready candidates exceed $\theta_{max}$, a fallback mechanism forces the substitution of the variable yielding the smallest absolute size $S_{min}$ (Lines 24-26). This penalty-aware approach gracefully navigates local rewriting, maintaining a concise polynomial size.

\begin{figure}[tbp]
    \centering
    \includegraphics[width=\linewidth]{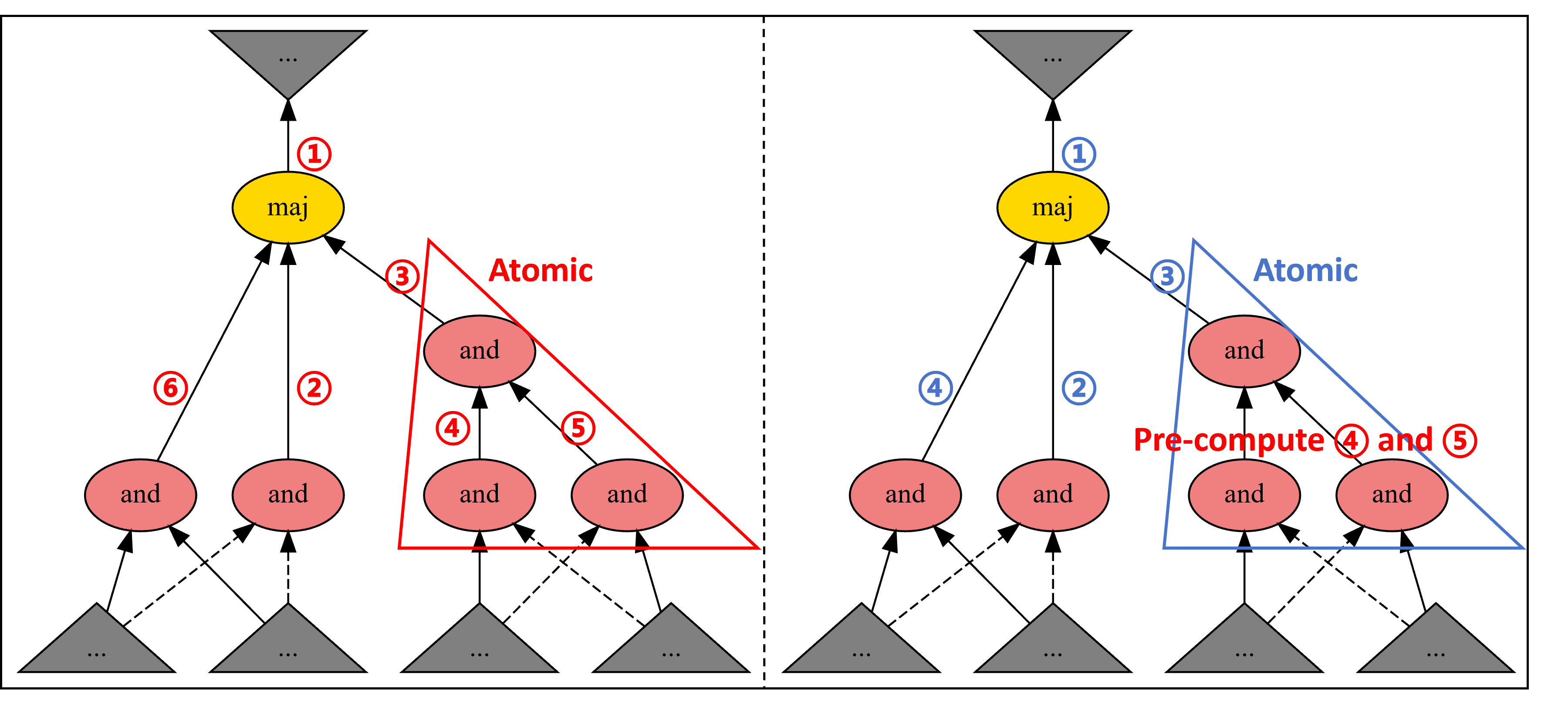}
    \caption{Two polynomial evaluation strategies for atomic blocks. Left: Localized rewriting of internal variables during the global reduction. Right: Direct pre-computation and caching of the root polynomial prior to the global reduction.}
    \label{fig:delay}
    \vspace{-2mm}
\end{figure}

To manage the reduction of internal variables of the atomic blocks detailed in Section \ref{subsec:atomic}, Arisca provides two evaluation strategies. These strategies are controlled through a configurable flag \texttt{delay}. As shown in the left part of Fig.~\ref{fig:delay}, enabling this flag triggers a localized rewriting phase. The global engine postpones the evaluation of atomic blocks and processes internal variables only after substituting the block's root variable in the main loop. The engine then temporarily enters a local context to reduce these internal variables. Shown in the right part, the alternative strategy performs direct pre-computation. Prior to global rewriting, the engine completely resolves and caches the root polynomial of each block. Since dynamic rewriting frequently reverts substitutions, cached polynomials prevent repeated evaluations during these failed trials. However, storing these polynomials requires a higher memory overhead. Therefore, the flag \texttt{delay} balances the memory footprint against computational efficiency.

\begin{algorithm}[tbp]
\caption{Dynamic Rewriting Ordering with Penalty-Aware Look-ahead}
\label{alg:ordering}
\begin{algorithmic}[1]
\renewcommand{\algorithmicrequire}{\textbf{Input:}}
\renewcommand{\algorithmicensure}{\textbf{Output:}}

\Require Target polynomial $\mathcal{P}$, Set of candidate variables $\mathcal{V}$, Explosion threshold $\theta_{max}$
\Ensure Fully rewritten polynomial in terms of Primary Inputs

\State Initialize penalty factors $p(v) \leftarrow 1$ for all $v \in \mathcal{V}$
\While{$\mathcal{V}$ is not empty}
    \State $\mathcal{V}_{ready} \leftarrow \{v \in \mathcal{V} \mid v \text{ is ready in topological order}\}$
    \For{each $v \in \mathcal{V}_{ready}$}
        \State $n_v \leftarrow \text{Number of occurrences of } v \text{ in } \mathcal{P}$
        \State $Score(v) \leftarrow p(v) \times (n_v \times |\mathcal{P}_v|) \text{ or } p(v)\times n_v$  \Comment{$|\mathcal{P}_v|$ is the size of the polynomial for variable $v$}
    \EndFor
    \State Sort $\mathcal{V}_{ready}$ in ascending order of $Score(v)$
    
    \State $accepted \leftarrow \text{False}$, $S_{min} \leftarrow \infty$, $v_{best} \leftarrow \text{null}$, $\mathcal{P}_{best} \leftarrow \text{null}$
    
    \For{each $v \in \mathcal{V}_{ready}$}
        \State $\mathcal{P}' \leftarrow \text{Substitute } v \text{ with } \mathcal{P}_v \text{ in } \mathcal{P}$
        \State $\Delta \leftarrow (|\mathcal{P}'| - |\mathcal{P}|) / |\mathcal{P}|$ \Comment{Calculate growth ratio}
        
        \If{$\Delta \le \theta_{max}$}
            \State $\mathcal{P} \leftarrow \mathcal{P}'$, $\mathcal{V} \leftarrow \mathcal{V} \setminus \{v\}$
            \State $accepted \leftarrow \text{True}$
            \State \textbf{break}
        \Else
            \State $p(v) \leftarrow p(v) \times 2$ \Comment{Penalize explosive variables}
            \If{$|\mathcal{P}'| < S_{min}$}
                \State $S_{min} \leftarrow |\mathcal{P}'|$, $v_{best} \leftarrow v$, $\mathcal{P}_{best} \leftarrow \mathcal{P}'$
            \EndIf
        \EndIf
    \EndFor
    
    \If{\textbf{not} $accepted$}
        \State $\mathcal{P} \leftarrow \mathcal{P}_{best}$, $\mathcal{V} \leftarrow \mathcal{V} \setminus \{v_{best}\}$
    \EndIf
\EndWhile

\State \textbf{return} $\mathcal{P}$
\end{algorithmic}
\end{algorithm}

\begin{algorithm}[tbp]
\caption{Greedy Variable Phase Inversion}
\label{alg:phase}
\begin{algorithmic}[1]
\renewcommand{\algorithmicrequire}{\textbf{Input:}}
\renewcommand{\algorithmicensure}{\textbf{Output:}}

\Require Current polynomial $\mathcal{P}$, Target variable $v$, Substitution polynomial $\mathcal{P}_v$, Set  of inverted variables $\mathcal{I}$
\Ensure Updated polynomial $\mathcal{P}$ and inverted set $\mathcal{I}$

\State $\mathcal{P}_v \leftarrow \text{PhaseCorrect}(\mathcal{P}_v, \mathcal{I})$ 
\State $\mathcal{P} \leftarrow \text{Substitute } v \text{ with } \mathcal{P}_v \text{ in } \mathcal{P}$

\State $\mathcal{V}_{new} \leftarrow \{u \mid u \text{ is introduced by } \mathcal{P}_v \text{ and } u \notin \mathcal{I}\}$
\For{each $u \in \mathcal{V}_{new}$}
    \State $\mathcal{P}' \leftarrow \text{Substitute } u \text{ with } 1 - u \text{ in } \mathcal{P}$
    \If{$|\mathcal{P}'| < |\mathcal{P}|$} 
        \State $\mathcal{P} \leftarrow \mathcal{P}', \quad \mathcal{I} \leftarrow \mathcal{I} \cup \{u\}$
    \EndIf
\EndFor

\State \textbf{return} $\mathcal{P}, \mathcal{I}$
\end{algorithmic}
\end{algorithm}

To further control the polynomial size, Arisca incorporates a greedy phase inversion technique \cite{kaufmann2022Adding, konrad2024Symbolic} into the dynamic rewriting engine. The Algorithm \ref{alg:phase} shows how to flip the polarity of intermediate variables by replacing $u$ with $1 - u$ to achieve a more compact representation. Initially, a global set $\mathcal{I}$ tracks all inverted variables. Before substituting a variable $v$, the engine aligns the polarities within $\mathcal{P}_v$ with $\mathcal{I}$. After the substitution, Arisca tentatively inverts the newly introduced variables. It commits this inversion and updates $\mathcal{I}$ only if the overall polynomial size strictly decreases. Otherwise, the engine reverses the flipping. This greedy heuristic inherently incurs computational overhead during the trial-and-error process. Furthermore, immediate local reductions do not always guarantee global optimality. To manage these tradeoffs, Arisca provides a configurable flag \texttt{flip} to selectively enable this feature.

\subsection{Portfolio}
\label{subsec:portfolio}

The aforementioned methodologies establish a highly configurable verification framework, essential for handling arithmetic circuits with diverse architectures. To address this structural sensitivity, Arisca implements a parallel portfolio strategy---concurrent workers execute multiple distinct configurations simultaneously. We frame this not as a brute-force workaround, but as a deterministic industrial asset for modern multi-core workstations. In commercial EDA environments, memory and compute time are much cheaper than the engineering hours wasted on debugging brittle, single-heuristic failures. To manage the inherent memory overhead of parallel rewriting, Arisca enforces a strict resource-bounding mechanism via a configurable threshold \texttt{size-limit}. The engine immediately terminates any worker whose polynomial exceeds this limit. This early termination promptly reclaims memory and CPU resources for more promising configurations. Section \ref{sec:exp} details the specific parameter combinations utilized in our portfolio.

\section{Evaluation}
\label{sec:exp}

\subsection{Experimental Setup}

Arisca is implemented in the Rust programming language. All experiments were conducted on an AMD Ryzen Threadripper PRO 9985WX CPU with 3.2GHz, with resources limited to 32GB of memory and $10^4$ seconds of time per instance. As reported in \cite{konrad2024Symbolic}, AMulet 2.2 \cite{kaufmann2023Improving} and TeluMA \cite{kaufmann2022Adding} exhibit comparable performance. Furthermore, our empirical analysis reveals that RevSCA-2.0 \cite{mahzoon2022RevSCA20} implicitly incorporates the heuristic proposed in \cite{mahzoon2020Formal}. Therefore, we select three representative baselines for our evaluation: AMulet 2.2, DynPhaseOrderOpt, and RevSCA-2.0 (hereafter referred to as AMulet, DynPOO, and RevSCA for brevity). 

The evaluated benchmark suite for multipliers is identical to that in \cite{konrad2024Symbolic}, comprising 310 diverse unsigned 64-bit multiplier circuits:
\begin{itemize}
    \item 192 multipliers from the Aoki benchmark \cite{homma2006Formal}.
    \item 28 multipliers generated by \texttt{GenMul} \cite{mahzoon2021GenMul}. 
    \item 90 multipliers generated by \texttt{multgen} \cite{temel2019Fast}.
\end{itemize}

To demonstrate Arisca's versatility beyond standard multipliers, we also evaluate its performance on general arithmetic circuits comprising addition and multiplication. Specifically, we generated various custom benchmarks using \texttt{multgen} as case studies. Table~\ref{tab:arithmetic} details the specifications and verification results for each circuit.

Arisca utilizes the default configuration ``\texttt{-m heuristic -r 5 -e adder {-}{-}max-ratio 0.01}''. Specifically, the \texttt{heuristic} mode enables the dynamic rewriting order, while \texttt{{-}{-}max-ratio} sets the expansion threshold $\theta_{max}$ defined in Algorithm \ref{alg:ordering}. All other optional flags are set to \texttt{false} by default. Within the parallel portfolio, individual workers override these baseline parameters with their specific assigned configurations.

\begin{table}[tbp]
    \centering
    \caption{Configurations and solved instances of the parallel portfolio workers in the original benchmark.}
    \label{tab:portfolio}
    \begin{tabular}{@{} c l c @{}}
        \toprule
        \textbf{ID} & \textbf{Best Arguments} & \textbf{\# of Solved Instances} \\
        \midrule
        0 & Default & 77 \\
        1 & \texttt{-r 9} & 82 \\
        2 & \texttt{--flip} & 10 \\
        3 & \texttt{--flip -r 10} & 20 \\
        4 & \texttt{--max-ratio 0.1 --flip -r 10} & 36 \\
        5 & \texttt{-m bfs} & 58 \\
        6 & \texttt{-m dfs} & 27 \\
        \bottomrule
    \end{tabular}
\end{table}
\begin{table}[tbp]
    \centering
    \caption{Overall performance and feature comparison of different tools.}
    \label{tab:tool_comparison}
    \begin{tabular}{@{} l c c c c c @{}}
        \toprule
        \multirow{2}{*}{\textbf{Tool}} & \multicolumn{3}{c}{\textbf{\# of Solved Instances}} & \multirow{2}{*}{\textbf{Open Source}} & \multirow{2}{*}{\textbf{Citation}} \\
        \cmidrule(lr){2-4}
        & \textbf{Original} & \textbf{\texttt{resyn3}} & \textbf{\texttt{dc2}} & & \\
        \midrule
        \textbf{Arisca} & \textbf{310} & \textbf{277} & \textbf{222} & \cmark & - \\
        AMulet          & 243 & 23  & 7   & \cmark & \cite{kaufmann2023Improving} \\
        DynPOO          & 299 & 272 & \textbf{222} & \xmark & \cite{konrad2024Symbolic} \\
        RevSCA          & 245 & 216 & 170 & \xmark & \cite{mahzoon2022RevSCA20} \\
        \bottomrule
    \end{tabular}
\end{table}
\begin{table}[tbp]
    \centering
    \caption{Verification performance of Arisca on diverse custom arithmetic circuits.}
    \label{tab:arithmetic}
    \begin{tabularx}{\linewidth}{@{} >{\raggedright\arraybackslash}X >{\raggedright\arraybackslash}X c @{}}
        \toprule
        \textbf{Circuit Description} & \textbf{Extra Arguments} & \textbf{Time} \\
        \midrule
        127$\times$129-bit signed multiplier with a 250-bit truncated output & ``\texttt{{-}{-}spec [127]*[129]} \texttt{{-}{-}signed}'' & 35.20 s \\ \addlinespace
        256-bit standard adder with a 1-bit carry-in                       & ``\texttt{{-}{-}spec [1]+[256]+} \texttt{[256]}''                & 15.34 ms \\ \addlinespace
        63$\times$65-bit MAC with a 128-bit addend           & ``\texttt{{-}{-}spec [63]*[65]+} \texttt{[128]}''                  & 2.44 s \\ \addlinespace
        Dot product of two vectors of length 16 with 8-bit elements          & ``\texttt{{-}{-}spec } \texttt{[8:0]*[8:128]+ $\cdots$ } \texttt{+[8:120]*[8:248]}'' & 310.37 ms \\
        \bottomrule
    \end{tabularx}
\end{table}
\subsection{Results}
As shown in Table~\ref{tab:portfolio}, seven workers are sufficient for Arisca to solve all 310 multipliers in the original benchmark. Specifically, the table records the number of instances in which each worker achieved the fastest verification time, even though other workers might also successfully solve the same cases. The table reveals that different parameter configurations excel in solving different multiplier architectures. This observation directly underscores the fundamental limitations of rigid heuristics and validates the necessity of our parameterized verification framework.

\begin{figure*}[tbp]
    \centering
    \begin{subfigure}[t]{0.3\textwidth}
        \vspace{0pt}
        \centering
        \includegraphics[width=\textwidth]{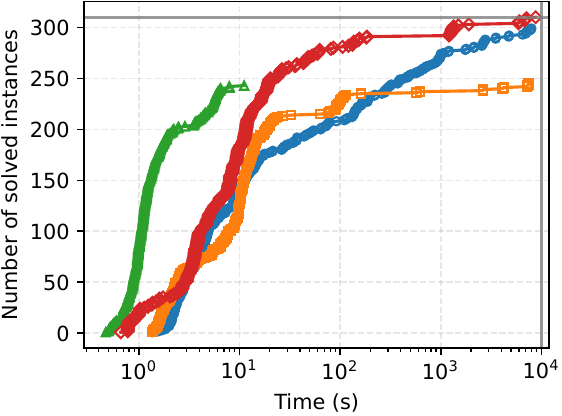}
        \caption{No Optimization.}
        \label{fig:cactus_simple_time}
    \end{subfigure}\hfill
    \begin{subfigure}[t]{0.3\textwidth}
        \vspace{0pt}
        \centering
        \includegraphics[width=\textwidth]{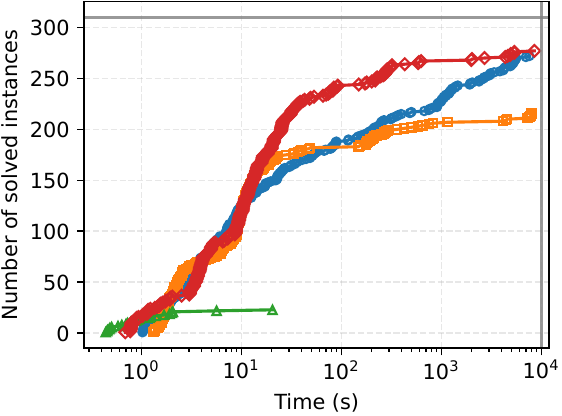}
        \caption{Optimized by \texttt{resyn3}.}
        \label{fig:cactus_dc2_time} 
    \end{subfigure}\hfill
    \begin{subfigure}[t]{0.3\textwidth}
        \vspace{0pt}
        \centering
        \includegraphics[width=\textwidth]{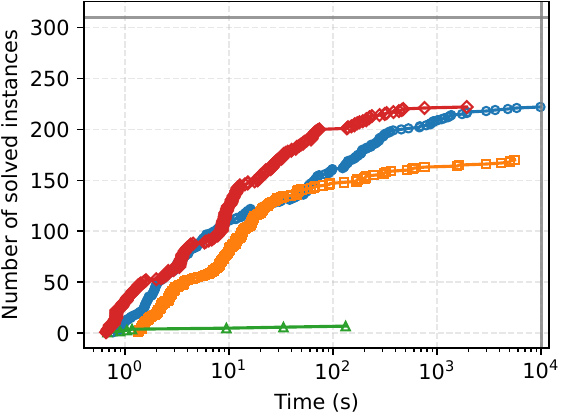}
        \caption{Optimized by \texttt{dc2}.}
        \label{fig:cactus_resyn3_time}
    \end{subfigure}\hfill
    \begin{minipage}[t]{0.09\textwidth}
        \vspace{0pt}
        \centering
        \includegraphics[width=\textwidth]{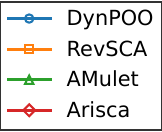}
    \end{minipage}
    \caption{The runtime performance of various verification tools among different benchmark optimizations.}
    \label{fig:cactus_time_all}
\end{figure*}

\begin{figure*}[tbp]
    \centering
    \begin{subfigure}[t]{0.3\textwidth}
        \vspace{0pt}
        \centering
        \includegraphics[width=\textwidth]{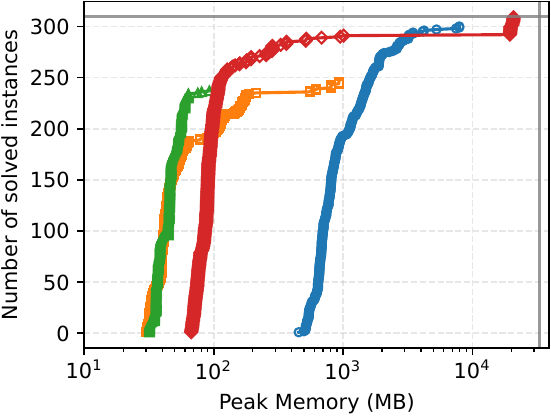}
        \caption{No Optimization.}
        \label{fig:cactus_simple_mem}
    \end{subfigure}\hfill
    \begin{subfigure}[t]{0.3\textwidth}
        \vspace{0pt}
        \centering
        \includegraphics[width=\textwidth]{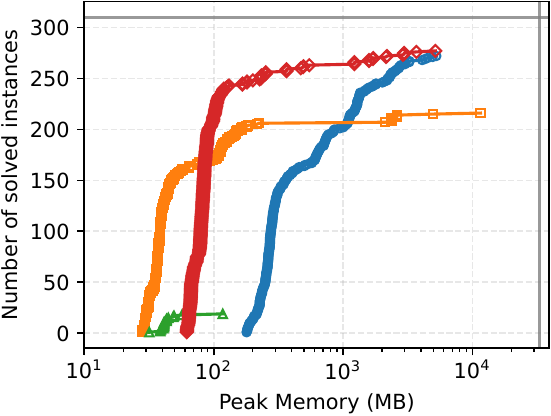}
        \caption{Optimized by \texttt{resyn3}.}
        \label{fig:cactus_dc2_mem} 
    \end{subfigure}\hfill
    \begin{subfigure}[t]{0.3\textwidth}
        \vspace{0pt}
        \centering
        \includegraphics[width=\textwidth]{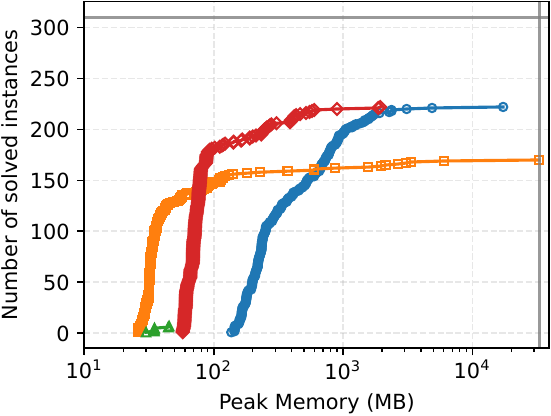}
        \caption{Optimized by \texttt{dc2}.}
        \label{fig:cactus_resyn3_mem}
    \end{subfigure}\hfill
    \begin{minipage}[t]{0.09\textwidth}
        \vspace{0pt}
        \centering
        \includegraphics[width=\textwidth]{fig/legend_only.pdf}
    \end{minipage}
    \caption{The peak memory of various verification tools among different benchmark optimizations.}
    \label{fig:cactus_memory_all}
\end{figure*}

In addition, as shown in Table~\ref{tab:tool_comparison}, Arisca is the only tool capable of solving all 310 instances. Arisca also achieves SOTA performance on benchmarks optimized by \texttt{dc2} and \texttt{resyn3}. In these optimized circuits, according to our experiments, inherently exclusive variables frequently entangle to form new vanishing monomials. Specifically, logic synthesis optimizations often introduce complex multi-fanout nodes. During backward rewriting, a conservative estimation strategy might trace and substitute along only one fanout path because it locally appears ``safer'', leaving the other fanout path completely unreduced. Consequently, the shared node and its successor variables end up multiplying together within the intermediate polynomials. As the product of these entangled variables might inherently evaluate to zero, their prolonged presence before cancellation causes a rapid accumulation of vanishing monomials. To resolve this, we introduce additional workers that enable the aggressive estimation heuristic introduced in Section \ref{subsec:order} and utilize larger \texttt{max-ratio} thresholds. This aggressive strategy encourages the engine to concurrently eliminate variables from multi-fanout paths, effectively disentangling exclusive variables and neutralizing the hidden vanishing monomials early in the reduction process.

Moreover, Fig.~\ref{fig:cactus_time_all} and Fig.~\ref{fig:cactus_memory_all} demonstrate Arisca's competitive runtime and memory efficiency. It should be clarified that the runtime plotted in Fig.~\ref{fig:cactus_time_all} corresponds to the elapsed CPU time of the fastest successful worker, rather than the cumulative execution time of all parallel threads. So does the peak memory plotted in Fig.~\ref{fig:cactus_memory_all}. Note that reporting the aggregate footprint of the entire portfolio is practically non-informative, as it simply reflects an arbitrary trade-off between memory-intensive parallelism and time-consuming serial execution. Instead, the winning worker highlights that our combined parameter space guarantees the existence of configurations that are both exceptionally fast and highly memory-efficient. Moreover, to safely manage the remaining concurrent overhead, Arisca enforces the strict resource-bounding mechanism discussed in Section \ref{subsec:portfolio}. By immediately terminating any thread that exceeds pre-defined intermediate sizes or memory thresholds, Arisca ensures that the cumulative memory footprint remains strictly within the capacity of modern workstations.

\begin{figure*}[tbp]
    \centering
    \hspace{0.05\linewidth}
    \begin{subfigure}[t]{0.28\textwidth}
        \vspace{0pt}
        \centering
        \includegraphics[width=\textwidth]{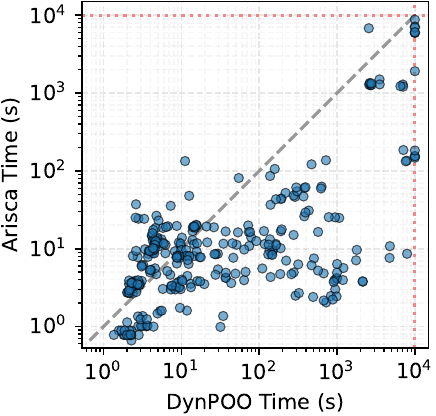}
        \caption{No Optimization.}
        \label{fig:scatter_simple}
    \end{subfigure}\hfill
    \begin{subfigure}[t]{0.28\textwidth}
        \vspace{0pt}
        \centering
        \includegraphics[width=\textwidth]{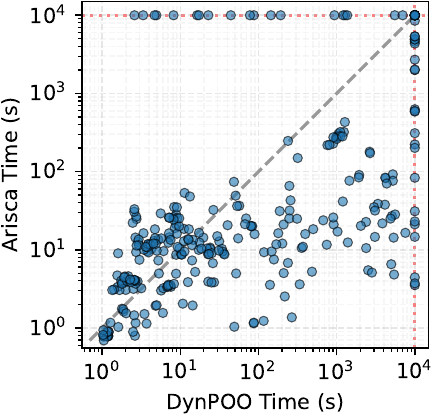}
        \caption{Optimized by \texttt{resyn3}.}
        \label{fig:scatter_resyn3}
    \end{subfigure}\hfill
    \begin{subfigure}[t]{0.28\textwidth}
        \vspace{0pt}
        \centering
        \includegraphics[width=\textwidth]{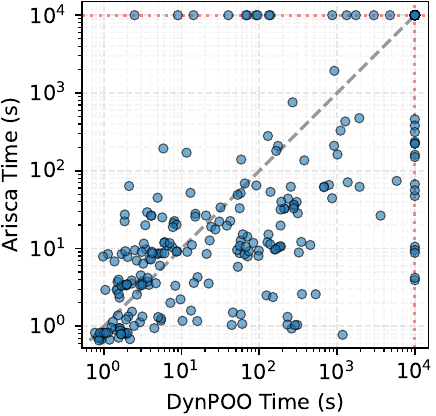}
        \caption{Optimized by \texttt{dc2}.}
        \label{fig:scatter_dc2}
    \end{subfigure}
    \hspace{0.05\linewidth}
    
    \caption{Comparing the time of DynPOO (X-axis) with Arisca (Y-axis).}
    \label{fig:scatter_all}
\end{figure*}
Among the baselines, the open-source AMulet demonstrates excellent runtime performance on simple multipliers. However, it struggles when final adder boundaries are blurry. In such cases, AMulet fails to replace the final adder and incurs significant runtime overhead. It even occasionally exhibits bugs that incorrectly substitute the final adder, corrupting the multiplier logic. DynPOO uses a robust fixed heuristic to achieve the previous SOTA performance. However, as illustrated in Fig.~\ref{fig:scatter_all}, Arisca achieves shorter verification times in the majority individual cases. As for RevSCA, its underlying heuristic is less robust compared to DynPOO. Furthermore, our empirical evaluation revealed that it identified certain correct multipliers as incorrect. This finding matches the experiment report of \cite{konrad2024Symbolic}. Our analysis indicates that it stems from its reverse engineering implementation, which does not strictly preserve logical equivalence during adder extraction and generates dangling nodes in the circuit. Conversely, both Arisca and DynPOO are bug free among these benchmarks. Finally, both DynPOO and RevSCA are closed-source. This lack of open availability severely hinders their broader industrial adoption.

As detailed in Table~\ref{tab:arithmetic}, we also evaluated Arisca on a diverse set of custom arithmetic circuits, including an asymmetric multiplier, an adder, a MAC and a dot-product unit. All instances are verified in just a few seconds. Although existing tools are strictly hardcoded to verify standard $n \times n$ multipliers, Arisca supports generic arithmetic circuits with any combination of addition and multiplication by implementing a general specification parser. 

\section{Related Work}
\label{sec:related}

In recent years, SCA has achieved remarkable success in formally verifying gate-level arithmetic circuits \cite{sayed-ahmed2016Formal, lv2013Efficient}. Building upon this foundational breakthrough, a variety of advanced algebraic reasoning frameworks have been proposed to tackle the scalability challenges of increasingly complex architectures of different arithmetic circuits.

To improve the verification scalability for large multipliers, Kaufmann et al. developed the open-source AMulet framework \cite{kaufmann2020Incremental, kaufmann2019Verifying, kaufmann2023Improving}. The early iterations proposed a hybrid methodology that structurally replaces the final adder of the multiplier architecture with a simple synthetic one to simplify backward rewriting, and then relied on SAT solvers to verify the replaced adder \cite{kaufmann2019Verifying}. Subsequently, the authors introduced XOR-based slicing to partition the circuit into distinct columns for structured reduction \cite{kaufmann2020Incremental}, which was later enhanced with improved data structures such as monomial sharing \cite{kaufmann2023Improving}. However, these structural techniques are heavily based on architectures preserving intact adder trees or clear XOR chains, which are frequently obfuscated in highly optimized multipliers. To bypass such structural dependencies, they introduced dual variables to maintain polynomial compactness by leveraging inverted signal phases \cite{kaufmann2022Adding}. Although theoretically inspiring, the algebraic computation required for dual variables introduced additional computational overhead, often negating practical performance benefits.

To tackle the polynomial explosion without relying on global macro-structures, Mahzoon et al. introduced the RevSCA series \cite{mahzoon2019RevSCA}. This approach demonstrates that trading minimal computational overhead for local reverse engineering to systematically extract HAs and FAs fundamentally simplifies the algebraic reduction by reducing intermediate variables. Based on this extracted knowledge, PolyCleaner \cite{mahzoon2018PolyCleaner} identifies that the structural reconvergence of HA output is a primary source of ``vanishing monomials''---a key factor that accounts for the explosion of polynomial size. To mitigate this, it leverages the sum-carry exclusion property (i.e., their product evaluating to zero) of the HA to proactively eliminate these redundant terms during the backward rewriting process. Subsequently, these local extraction and vanishing removal techniques were consolidated into a single robust flow in RevSCA-2.0 \cite{mahzoon2022RevSCA20}.

In addition to structural optimizations, the order of elimination of variables significantly impacts the verification performance. Since static topological rewriting often triggers intractable polynomial explosion, Mahzoon et al. introduced a dynamic polynomial substitution strategy \cite{mahzoon2020Formal}. This method evaluates tentative pre-eliminations and revokes substitutions that exceed a predefined size threshold. Subsequently, Konrad et al. extended this dynamic ordering paradigm by integrating an additional phase optimization heuristic \cite{konrad2024Symbolic}. Unlike \cite{kaufmann2022Adding}, this strategy permanently inverts a variable's polarity as long as it reduces the polynomial size.

Currently, the scope of the frameworks mentioned above remains restricted to standard multipliers where the two inputs have the same bit-width. However, their core algebraic principles can be directly ported to other fundamental components, such as adders and MAC units, without requiring ad-hoc optimizations. Furthermore, recent studies have demonstrated the viability of SCA in verifying more intricate designs, including finite field multipliers\cite{lv2013Efficient}, modular multipliers \cite{mahzoon2022Formal}, and integer dividers \cite{drechsler2023Divide, konrad2025Divider}. However, despite these isolated extensions, the community still lacks an open-source and generalizable framework capable of supporting this diverse spectrum of arithmetic circuits.

\section{Conclusion}
\label{sec:conclusion}

In this paper, we introduce Arisca, a highly adaptable verification engine built on the Symbolic Computer Algebra framework. Unlike existing specialized tools, Arisca extends its formal verification capabilities beyond standard multipliers to include general arithmetic components that comprise addition and multiplication. By systematically integrating, upgrading, and parameterizing previously proven but rigidly fixed SCA heuristics, Arisca achieves adaptability across diverse and complex circuit architectures. Extensive evaluations demonstrate that Arisca not only achieves SOTA performance on a comprehensive suite of multiplier benchmarks and their heavily optimized variants, but also exhibits good efficiency in verifying custom arithmetic components, such as MACs and dot-product units.


{
\clearpage
\balance
\bibliographystyle{ACM-Reference-Format}
\bibliography{./references}
}

\end{document}